# Scheme and Scale Dependence of Charm Production in Neutrino Scattering

G. Kramer[a], B. Lampe[b], H. Spiesberger[c]

[a]II. Institut für Theoretische Physik[*],[†]
Universität Hamburg
D - 22761 Hamburg, Germany

[b]Max-Planck-Institut für Physik, Werner-Heisenberg- Institut
D - 80805 München, Germany

[c]Fakultät für Physik[†]
Universität Bielefeld
D - 33615 Bielefeld, Germany

hep-ph/9511396    23 NOV 95

## Abstract

We discuss some theoretical uncertainties in the calculation of the cross section for charm production in charged current deep inelastic neutrino scattering related to ambiguities in the treatment of terms which are singular in the limit of a vanishing charm mass. In particular we compare the so-called variable flavour scheme where these terms are absorbed in the parton distribution functions containing the charm as an active flavour, with the so-called fixed flavour scheme with no charm mass subtraction where the charm appears only in the final state of fixed-order scattering matrix elements. Using available parametrizations of parton distribution functions we find that the two schemes lead to largely differing results for separate structure functions whereas the differences cancel to a large extent in the total cross section in that kinematical region which has been measured so far.

[*] Supported by Bundesministerium für Forschung und Technologie, Bonn, Germany, Contract 05 6HH93P(5).
[†] Supported by the EEC Program 'Human Capital and Mobility' through Network 'Physics at High Energy Colliders' under Contract CHRX-CT93-0357 (DG 12 COMA).

# 1 Introduction

Information on parton distributions comes mostly from totally inclusive lepton scattering on nucleons. The increased precision in the experimental measurements of deep inelastic scattering (DIS) and related processes has lead to a considerable improvement in our knowledge of the parton distributions of the proton and the neutron. Various global analyses of the data have been done in recent years to obtain increasingly better parton densities [1, 2, 3]. All these analyses, however, give only limited information on the densities of heavy quarks, strange, charm and bottom, since they make a much smaller contribution to the measured structure functions.

This situation has improved recently for the strange-sea density. To extract this density from deep inelastic data two approaches have been used. In the first method one considers the difference of the $F_2$ structure functions measured in neutrino and muon DIS which selects the strange-sea contribution [4]. This method suffers from experimental normalization uncertainties since data from different experiments are combined. In this paper we are interested in the second method which consists in measuring the charm production in charged current neutrino and antineutrino DIS where the characteristic signature is the production of dimuons in the final state [5]. The dimuon data obtained by the CCFR collaboration were originally analyzed in the simple parton model as scattering of strange quarks in the nucleon supplemented with the slow-rescaling prescription to account for the effect of the finite mass of the charm quark [6]. The result of this analysis in leading order QCD (LO) was a strange-sea contribution with a strength reduced by a factor of $\simeq 0.4$ as compared to the non-strange quark and antiquark components of the sea and a fitted charm quark mass $m = 1.31\ GeV$. Later this analysis was repeated using the next-to-leading order (NLO) formalism of Aivazis et al. [7] with the result that the strange-sea distribution was still reduced by a factor of 0.5 compared to the non-strange-sea distribution, but the charm quark mass was now changed to $m = 1.70\ GeV$. This 50 % suppression in relation to the $u$ and $d$ sea distributions is now incorporated in all recent parton distribution fits [1, 2, 3]. Thus the $s$ distributions in MRS(A) [1] and CTEQ3M [2] agree quite well with the NLO strange-sea distributions of the CCFR collaboration [5] whereas in the recent GRV analysis [3] the generated $s$ distribution matches the LO distribution in the earlier CCFR analysis [6].

It is well known that in NLO the parton distribution are ambiguous and have no direct physical meaning since they depend on the subtraction or factorization scheme. Thus, in order that the $s$ distribution assumed in the MRS or the most recent CTEQ analyses are consistent with the CCFR results, the same scheme has to be chosen. The CCFR analysis is based on the NLO formalism of Aivazis et al. [7]. Their calculation of the charm production cross section is performed in the $\overline{MS}$ scheme with subtraction of the collinear singularity $m \to 0$. This is the variable flavour scheme in the terminology of ref. [7]. This scheme is preferred for situations with $Q^2 \gg m^2$. The factorization scale in [7] is chosen to be $\mu = 2p_T^{max}$ where $p_T^{max} = \Delta(W^2, m^2, M^2)/2W$ is the maximum available transverse momentum of the final state charm quark for given kinematic variables $x$ and $Q^2$. $M$ is the nucleon mass and $W$ the total c.m. energy. This scale $\mu$ is somewhat larger than $Q$ depending on the relevant value of $x$ which is $\langle x_{vis} \rangle = 0.15$ in the CCFR analysis. Except for this special choice of the factorization scale, which is also used for the scale of the $QCD$ coupling constant, the variable flavour scheme corresponds to the subtraction applied in the MRS and CTEQ analyses. This prescription implies that the charm contribution is introduced as an active flavour of the nucleon which absorbs the charm mass collinear singularity in the NLO current-gluon fusion contribution. This also implies that the leading logarithms due to the charm mass singularity are resummed through the usual QCD evolution of the charm distribution. Actually, in the MRS and CTEQ analyses the charm mass is equal to zero except for the thresholds at $Q = m$ introduced in $\alpha_s(Q^2)$ and in the charm parton distribution. The GRV collaboration, on the other hand, in a recent structure function analysis [3], have taken the point of view that the charm contribution in the nucleon can be generated dynamically through perturbative contributions and with no extra non-perturbative charm contribution that would have to be determined from DIS and other experimental data. This is the so-called fixed flavour scheme with $N_f = 3$ in the terminology of [7]. Here the structure functions $F_2^c$ and $F_L^c$ in the neutral current process are provided by fixed-order perturbation theory from the LO fusion process $\gamma^* g \to c\bar{c}$. The scale $\mu$ in $\alpha_s(\mu^2)$ and the gluon distribution function is chosen as $\mu = 2m$, irrespective of $Q^2$, which according to [8] leads to results similar to the corresponding NLO results of Laenen et al. [9]. A comparison of the fixed and variable flavour schemes has been performed for neutral current heavy quark production in [10].

Actually in this GRV analysis [3] the strange sea distribution is also generated dynamically, and the strange sea at their input scale $\mu_0^2$, which is very small, vanishes. It is clear that in the fixed flavour scheme, where the charm quark is totally extrinsic, the charm production in charged current processes, when considered in NLO, must be calculated also in the same scheme. Such a calculation has been performed some time ago by two of us [11] with the intention to study the charm mass dependence in relation to the higher order QCD corrections. At that time no attempt was made to analyze the scheme dependence and to compare the fixed flavour scheme with $N_f = 3$ with the variable flavour scheme with $N_f = 4$. Since now these two schemes have been applied in the construction of parton distributions for the nucleon, it is appropriate to find out, whether in both of these two schemes the dimuon production cross sections in charged current neutrino and antineutrino DIS can be obtained in NLO in agreement with experiment. Since the measurements are performed in an intermediate $Q^2$ range ($\simeq \langle Q_{vis}^2 \rangle = 25.5 \ GeV^2$ [5]) there is a priori no reason to prefer the variable flavour scheme as was done in the analysis of the experimental data [5]. Thus it would be advisable to repeat the experimental analysis with the fixed flavour scheme to obtain the NLO strange sea distribution in a consistent way in this scheme, too. Due to the experimental details involved in such an analysis this could be done only by the CCFR collaboration. Instead we take the following approach. The cross section for charm production depends in general on three independent structure functions $F_1^{sc}$, $F_2^{sc}$ and $F_3^{sc}$. We calculated these structure functions in the variable flavour scheme with the same input as in the CCFR analysis and consider the result as a representation of the experimental results, of course only in the range $0.01 < x < 0.2$, as measured in the CCFR experiment. For the proton (neutron) structure functions we use the CTEQ3M or the MRS(A) parton distribution functions which agree with the CCFR NLO strange quark distribution function up to small deviations. This can be done easily by converting the formalism in [11], which was obtained for unsubtracted charm, to the variable flavour scheme, i.e. with intrinsic charm. Then we calculated the $F_1^{sc}$, $F_2^{sc}$ and $F_3^{sc}$ structure functions in the fixed flavour scheme using the parton distribution functions of GRV [3], which are the only available parametrizations of this scheme, and compared the results in the two schemes. In addition we have investigated the effect of scale variations, in particular by replacing the scale by $Q^2$, which might be the more natural choice instead of $(p_T^{max})^2$.

The outline of this work is as follows. In section 2 we repeat some necessary formalism and specify our input. The numerical results for the charged current structure functions $F_1^{sc}$, $F_2^{sc}$ and $F_3^{sc}$ in the two schemes are presented and compared in section 3. Here we also discuss to what extent differences for the structure functions obtained for the two schemes might have shown up in the CCFR data, since in the CCFR experiment only some particular average over the three structure functions could be measured.

## 2 Formalism and Input

The signature for the production of charm quarks in neutrino- and antineutrino-nucleon scattering is the presence of two oppositely charged muons in the final state. For neutrino scattering on protons the underlying process is a neutrino interacting with a $s$ or a $d$ quark, producing a charm quark that fragments into a charmed hadron. The charmed hadron's semileptonic decay produces a second muon of a charge opposite to that of the first one. The analogous reaction with an ingoing antineutrino proceeds through an interaction with a $\bar{s}$ or $\bar{d}$ antiquark, again leading to oppositely charged muons in the final state.

The inclusive cross section for charm production in $\nu_\mu N$ collisions ($N$ stands for proton or neutron) $\nu_\mu N \to \mu^- c X$ is described by the structure functions $F_1^{sc}$, $F_2^{sc}$ and $F_3^{sc}$, which depend on $x$ and $Q^2$, and is given by the following formula:

$$\frac{d^2\sigma}{dxdy} = \frac{G_F^2}{2\pi} S \left( (1-y) F_2^{sc}(x, Q^2) + y^2 x F_1^{sc}(x, Q^2) + y(1-y/2) x F_3^{sc}(x, Q^2) \right). \tag{1}$$

For the reaction $\bar{\nu}_\mu N \to \mu^+ \bar{c} X$ the structure functions $F_1^{sc}, F_2^{sc}$ and $F_3^{sc}$ are replaced by $\overline{F_1^{sc}}, \overline{F_2^{sc}}$ and $(-\overline{F_3^{sc}})$. $G_F$ is the Fermi coupling and $x = Q^2/2Pq$, $S = (k+P)^2$, $y = Pq/Pk = Q^2/xS$ with $P$ and $k$ being the momenta of the ingoing nucleon and neutrino, respectively. The effect of the $W$ propagator is neglected as usual.

At leading order (LO), charm is produced by scattering directly from $s$ and $d$ quarks in the proton ($s$ and $u$ quarks in the neutron). For scattering off an isoscalar target the cross section is given by (1) where the structure functions in LO are given by:

$$F_2^{sc}(\bar{x}, Q^2) = \bar{x} \left( |V_{cd}|^2 (u(\bar{x}, Q^2) + d(\bar{x}, Q^2)) + |V_{cs}|^2 \, 2s(\bar{x}, Q^2) \right) = 2\bar{x} s'(\bar{x}, Q^2)$$

$$F_1^{sc}(\bar{x}, Q^2) = \frac{1}{2} F_3^{sc}(\bar{x}, Q^2) = s'(\bar{x}, Q^2) \tag{2}$$

where $\bar{x}$ is the slow rescaling variable, which takes into account the non-vanishing charm mass $m$, $\bar{x} = (1 + m^2/Q^2)x$ [12]. $V_{cd}$ and $V_{cs}$ are the CKM mixing matrix elements and $s'(x)$ is the 'effective' strange sea distribution in LO. The LO expression, given by (1) and (2), shows the sensitivity of the process to the strange quark sea distribution. Charm (antivcharm) production from scattering off $d$ ($\bar{d}$) quarks, respectively $u$ ($\bar{u}$) quarks, is Cabibbo suppressed. In the case of charm produced by neutrinos, half of the cross section is due to scattering from $s$ quarks, although the $d$ quark content of the proton is ten times larger. In antineutrino scattering, where $\bar{d}$ ($\bar{u}$) quarks from the sea contribute, approximately 90 % of the cross section is due to scattering of $\bar{s}$ quarks. In next-to-leading order, we have contributions proportional to $O(\alpha_s)$ which come from virtual corrections to the LO diagrams, gluon emission contributions and contributions from $W$-gluon fusion. The LO and NLO hard scattering diagrams are shown in Fig. 1. Due to the size of the gluon distribution, which is an order of magnitude larger than the sea quark distributions, the gluon-initiated diagram is of similar magnitude as the LO contribution. The NLO quark-initiated diagrams in which a gluon is radiated are smaller, since they are not enhanced by a large parton distribution.

In our earlier work [11] (see also [13]) we have calculated the $O(\alpha_s)$ corrections to the structure functions $F_1^{sc}, F_2^{sc}$ and $F_3^{sc}$ for a finite $c$ quark mass $m$ on the basis of the diagrams in Fig. 1 using dimensional regularization. The formulas given in [11] are for the fixed flavour scheme with the DIS convention for the subtraction of the collinear singularity of the intermediate $s$ quark. In this scheme the $F_3^{sc}$ structure function, for example, was given by (see (2.3b) in [11])

$$F_3^{sc}(\bar{x}, Q^2) = s'(\bar{x}, Q^2) + \frac{\alpha_s(\mu_R^2)}{2\pi} \int_{\bar{x}}^1 \frac{d\bar{z}}{\bar{z}} \left( \left[ d_{qq}\left(\bar{z}, \frac{m^2}{Q^2}\right) - f_{qq}(\bar{z}) \right] s'\left(\frac{\bar{x}}{\bar{z}}, Q^2\right) \right.$$
$$\left. + \left[ d_{qg}\left(\bar{z}, \frac{m^2}{Q^2}\right) - f_{qg}(\bar{z}) \right] g\left(\frac{\bar{x}}{\bar{z}}, Q^2\right) \right). \quad (3)$$

In the following we shall define the various subtraction schemes for the contribution induced by the gluon in the nucleon, i.e. by the process $W^+ g \to c\bar{s}$ which is the term proportional to $g(\bar{x}/\bar{z}, Q^2)$ in (3).

In the scheme with no charm subtraction the functions $d_{qg}(\bar{z}, \frac{m^2}{Q^2})$ and $f_{qg}(\bar{z})$ have the following form:

$$d_{qg} = T_R \left( (1-\bar{z})^2 + \bar{z}^2 + 4z\frac{m^2}{Q^2}(1-\bar{z}) - \left[ (1-\bar{z})^2 + \bar{z}^2 + 4z\frac{m^2}{Q^2}\left(1 - 2z - z\frac{m^2}{Q^2}\right) \right] \ln \frac{s}{m^2} \right.$$
$$\left. + [(1-\bar{z})^2 + \bar{z}^2] \ln \frac{(1-\bar{z})^2}{z(1-z)} \right), \quad (4)$$

$$f_{qg}(\bar{z}) = T_R \left( [(1-\bar{z})^2 + \bar{z}^2] \left( \ln \frac{1-\bar{z}}{\bar{z}} + \ln \frac{\mu_s^2}{Q^2} \right) + 8\bar{z}(1-\bar{z}) - 1 \right) \quad (5)$$

where $s = Q^2(1-z)/z$ and $T_R = 1/2$. In (4) we must distinguish $z$ and $\bar{z} = (1+m^2/Q^2)z$. In [11] the factorization scale for the $s$ quark was chosen as $\mu_s^2 = Q^2$. If we have an arbitrary scale, it appears as an additional term in (5). The expression for $f_{qg}(\bar{z})$ in (5) corresponds to the DIS scheme. In (5) we corrected an error [14] which is already present in the original calculation [15]. The $f_{qg}$ given in [11, 15] is not exactly in the $\overline{MS}$ scheme because initial gluon spins were averaged in four rather than in $n$ dimensions. The form as given in (5) must be used for the case that the neutrino cross sections are calculated with nucleon structure functions constructed in the DIS scheme. When using structure functions in the $\overline{MS}$ scheme, (5) is replaced by the simpler expression

$$f_{qg}(\bar{z}) = T_R \left( (1-\bar{z})^2 + \bar{z}^2 \right) \ln \frac{\mu_s^2}{Q^2}. \quad (6)$$

In the variable flavour scheme, the leading singularity in the charm mass $\propto \ln m^2$ is subtracted as well. For the $\overline{MS}$ scheme, this subtraction is very simple. We combine it with $d_{qg}$, which then contains all mass dependent terms. Then the total $d_{qg}^{\overline{MS}}$ is

$$d_{qg}^{\overline{MS}} = d_{qg} + \Delta d_{qg}^{\overline{MS}} \tag{7}$$

with

$$\Delta d_{qg}^{\overline{MS}} = T_R \left((1-\bar{z})^2 + \bar{z}^2\right) \ln \frac{\mu_s^2}{m^2} \tag{8}$$

where $\mu_s$ is the subtraction scale for the charm term. We see that the term $\Delta d_{qg}^{\overline{MS}}$ subtracts just the contribution proportional to $\ln m^2$ in (4), and the total $d_{qg}^{\overline{MS}}$ has no singularity for $m \to 0$.

In the DIS scheme we must subtract additional terms which are equal to the contribution to the $F_2$ structure function in neutral current deep inelastic scattering originating from the transition $\gamma g \to c\bar{c}$. This subtraction is not unique. In our calculation we keep the charm mass $m$ finite whereas in all structure function analyses (except those of GRV) the $c$ mass is neglected except for a threshold in the $Q^2$ evolution related to $m$. From this viewpoint it would be reasonable to subtract terms corresponding to the massless cross section for $\gamma g \to c\bar{c}$ except for the term proportional to $\ln(\mu_s^2/m^2)$. In this case the subtraction term has the same structure as $f_{qg}$ in (5) and $d_{qg}$ has to be modified by an additional term $\Delta d_{qg}^{DIS}$ which is

$$\Delta d_{qg}^{DIS} = T_R \left([(1-\bar{z})^2 + \bar{z}^2]\left(\ln \frac{1-\bar{z}}{\bar{z}} + \ln \frac{\mu_s^2}{m^2}\right) + 8\bar{z}(1-\bar{z}) - 1\right). \tag{9}$$

Of course, (8) still contains mass dependent terms since $z$ is replaced by $\bar{z}$, i.e. we use the slow-rescaling form of the variable $z$ as in $f_{qg}$. This is partially dictated by the requirement that the terms proportional to $\ln m^2$ must cancel [13]. We observe that with this prescription terms proportional to $m^2 \ln m^2$ survive.

Another possibility is to follow Gottschalk [13] and to subtract the contribution of $F_2$ for massive quarks which is obtained from $\gamma g \to c\bar{c}$ with $m \neq 0$. Here we have the problem that the thresholds for $\gamma g \to c\bar{c}$ and $W^+ g \to c\bar{s}$ are different, $4m^2$ for the first process and $m^2$ for the second process. The subtraction in [13], which we call the $DIS_m$ scheme, proceeds as follows. The cross section for $\gamma g \to c\bar{c}$ is given by $c_{g,2}(z, \frac{m^2}{Q^2})$ (this is the term equivalent to $d_{qg}(z, \frac{m^2}{Q^2})$ in (3)) [16]:

$$c_{g,2}^c(z, \frac{m^2}{Q^2}) = T_R \left(\left[(1-z)^2 + z^2 + z(1-3z)\frac{4m^2}{Q^2} - z^2\frac{8m^4}{Q^4}\right] \ln \frac{1+\beta}{1-\beta} \right.$$
$$\left. + \beta \left[8z(1-z) - 1 - z(1-z)\frac{4m^2}{Q^2}\right]\right). \tag{10}$$

In (10), $z$ is the variable with no slow-rescaling factor, i.e. $z = Q^2/(Q^2 + s)$ and

$$\beta = \left(1 - \frac{4m^2}{Q^2}\frac{z}{1-z}\right)^{1/2} = \left(1 - \frac{4m^2}{s}\right)^{1/2}. \tag{11}$$

The slow-rescaling variable for $\gamma g \to c\bar{c}$ is $\bar{\bar{z}} = (Q^2 + 4m^2)/(Q^2 + s) = (1 + 4m^2/Q^2)z$, which has the property, that for $\bar{\bar{z}} \to 1$ we have $s \to 4m^2$, so that the argument of the square root in (11) is always positive. This is not the case if we would express $z$ by $\bar{z}$ and would consider $\bar{z} \to 1$. The subtraction term in the $DIS_m$ scheme can be written as:

$$\Delta d_{qg}^{DIS_m} = c_{g,2}^c\left(z, \frac{m^2}{Q^2}\right)\frac{z}{\bar{\bar{z}}} + T_R[(1 - \bar{\bar{z}})^2 + \bar{\bar{z}}^2]\ln\frac{\mu_s^2}{Q^2} \tag{12}$$

where $z$ must be expressed by the variable $\bar{\bar{z}}$ using the relation given above. In a second step $\bar{\bar{z}}$ is identified with the slow-rescaling variable $\bar{z}$ from the process $W^+g \to c\bar{s}$. This procedure is certainly not unique. (12) differs from (8) by terms proportional to $m^2/Q^2$. For $Q^2 \gg m^2$ these two ways of subtraction give approximately the same result.

We shall not write down the equivalent subtraction terms for the contributions induced by the $s$ quark in (3), i.e. by the process $W^+s \to cg$. In the scheme with no charm subtraction the function $d_{qq}(\bar{z}, \frac{m^2}{Q^2})$ was given in [11] together with the subtraction term $f_{qq}(\bar{z})$ in the $DIS$ scheme for $\mu_s^2 = Q^2$. The form of this term for $\mu_s^2 \neq Q^2$ is obvious from (5) and the $\overline{MS}$ subtraction term is analogous to (6). The charm subtraction terms $\Delta d_{qq}^{\overline{MS}}$, $\Delta d_{qq}^{DIS}$ and $\Delta d_{qq}^{DIS_m}$ are constructed analogously to (7), (9) and (12). In the same way one proceeds for the structure functions $F_1^{sc}(\bar{x}, Q^2)$ and $F_2^{sc}(\bar{x}, Q^2)$. For these the $O(\alpha_s)$ corrections have been given also in [11] and the various subtraction terms are constructed in the same way as for $F_3^{sc}(\bar{x}, Q^2)$.

The calculation of the structure functions $F_1^{sc}(\bar{x}, Q^2)$, $F_2^{sc}(\bar{x}, Q^2)$, $F_3^{sc}(\bar{x}, Q^2)$ and

$$F_L^{sc}(\bar{x}, Q^2) = F_2^{sc}(\bar{x}, Q^2) - 2x\,F_1sc(\bar{x}, Q^2) \tag{13}$$

for charm production by neutrinos in the various schemes has been done with the following nucleon structure function sets. For the variable flavour scheme, where the charm is treated as intrinsic, we have chosen the sets CTEQ3 [2] and MRS(A) [1] both in the $\overline{MS}$ and $DIS$ schemes. For these we have chosen the scale $\mu_R = \mu_s = 2p_T^{max}$ and $m = 1.7\,GeV$ as in the CCFR analysis [5]. In the following we shall refer to this scheme also as the $N_f = 4$ scheme. In the fixed flavour scheme, also denoted as the $N_f = 3$ scheme, we work with the parton distributions GRV(94) [3]. In this new structure function analysis the charm quark distribution to the electromagnetic structure functions $F_1$ and $F_2$ is generated dynamically from $\gamma^*g \to c\bar{c}$ at the scale $\mu_R = \mu_s = 2m$ with $m = 1.5\,GeV$. These parton distributions exist also in a $\overline{MS}$ and a $DIS$ version. To obtain the structure functions $F_i^{sc}(\bar{x}, Q^2)$ as they appear in the CCFR analysis we have calculated them for an isoscalar target including the Cabibbo suppressed terms of the charged current with $|V_{cd}| = 0.224$. This value is to be compared with the PDG value $V_{us} = 0.2205 \pm 0.0018$ and with $|V_{cd}| = 0.232 + 0.018(-0.020)$ as determined in the CCFR analysis [5], furthermore $|V_{cs}|^2 = 1 - |V_{cd}|^2$.

In the next section we present our numerical results and compare the charm production structure functions for CTEQ, MRS(A) and GRV with each other for the $\overline{MS}$ and $DIS$ schemes.

## 3 Numerical Results

Before we come to the charm production structure functions, we compare the most important input, i.e. the $s$ quark parton distribution of the proton. It is plotted at $\mu^2 = 25\,GeV^2$ for CTEQ3M and MRS(A) in the $\overline{MS}$ version and GRV(94) also in the $\overline{MS}$ version as a function of $x$ between $x = 0.01$ and 1 in Fig. 2 including the Cabibbo suppressed contribution coming from $u$ and $d$ quarks for the case of an isoscalar target, which is (see (2)):

$$s'(x,\mu^2) = |V_{cs}|^2 s(x,\mu^2) + |V_{cd}|^2 \frac{1}{2}(u(x,\mu^2) + d(x,\mu^2)). \tag{14}$$

We see that the CTEQ3 and MRS(A) distributions are almost equal in the considered range of $x$ whereas the $s'$ distribution of GRV(94) differs from the other two. It is clear that this difference comes from $s(x)$. The small admixture from $u(x)$ and $d(x)$ in (13) has no effect since the valence parts of these quark distributions are equal to those of MRS(A) by construction [3]. Since GRV use a different scheme for the charm quark, there need not be any agreement for the $s$ quark either since the distribution functions in NLO have no physical meaning. The only requirement is that the usual deep inelastic and other data are reproduced. We also mention that the strange quark distribution $xs(x)$ in MRS(A) [1] has been compared to the distribution obtained in the CCFR analysis [5]. At $\mu^2 = 4\ GeV^2$ it lies inside the error band for $xs(x)$ as given by CCFR and so satisfies the experimental constraint of reproducing the dimuon data if calculated with the same NLO formalism.

Next we present the comparison of results for the structure functions in the $\overline{MS}$ scheme. We choose $Q^2 = 25\ GeV^2$ since this corresponds very well to the average $\langle Q^2 \rangle = 25.5\ GeV^2$ in the CCFR analysis. Figs. 3, 4, 5 and 6 show the plots for $xF_1^{sc}(x,Q^2), F_2^{sc}(x,Q^2), xF_3^{sc}(x,Q^2)$ and $F_L^{sc}(x,Q^2)$ as a function of $x$ in the interval $[0.01, 1]$. We observe that for all four structure functions there is little difference between the CTEQ3 and MRS(A) results as to be expected from the comparison in Fig. 2 (note that also the gluon densities of the CTEQ3 and MRS(A) parametrizations are very similar). If we compare with the LO approximation which is essentially given in Fig. 2, up to a factor of 2 for $F_2^{sc}$ and $xF_3^{sc}$, we see that for CTEQ3 and MRS(A) the $O(\alpha_s)$ corrections together with the subtraction terms change $xF_1^{sc}, F_2^{sc}$ and $xF_3^{sc}$, in particular towards small $x$. $xF_1^{sc}$ and $F_2^{sc}$ have decreased and $xF_3^{sc}$ has increased strongly. In the GRV scheme, $xF_1^{sc}$ and $F_2^{sc}$ increased as compared to $xs'(x)$, but $xF_3^{sc}$ now decreased strongly. One can understand this pattern by looking at the different subtraction terms for the two cases which were given in the last section. The dominant subtraction term for the $N_f = 4$ structure functions (CTEQ3 and MRS(A)) contains the factor $\ln(\mu_s^2/m^2)$ with $\mu_s^2 = 4(p_T^{max})^2$, $p_T^{max} = (W^2 - m^2)/2W$, and rises like $\ln(1/x)$ for decreasing $x$, whereas the equivalent factor for the GRV distribution, $\ln(Q^2/m^2)$, is independent of $x$. Thus, the increase of the GRV strange sea towards small $x$ is reflected unchanged in the results for the structure functions, whereas the less steep strange sea of the CTEQ3 and MRS(A) distributions is made even less steep. We emphasize that the scales in the $N_f = 4$ and in the $N_f = 3$ case are different. For $x > 0.1$ the functions $xF_1^{sc}$ and $F_2^{sc}$ are almost equal in the two schemes but they differ appreciably for $x < 0.1$, whereas $xF_3^{sc}$ is much smaller in the $N_f = 3$ scheme for all $x$. $F_L^{sc}$ shown in Fig. 6 differs also in the two schemes, but now more for $x > 0.05$. However $F_L^{sc}$ is small and certainly very difficult to be measured.

In case that the three structure functions could be determined experimentally by measuring the charm production cross section in (1) as a function of $y$ with fixed $x$ and $Q^2$, the behaviour of these structure functions as a function of $x$, as shown in Figs. 3, 4 and 5, could be verified. It is clear that the structure functions in the two schemes are not compatible with each other, so that either one of them or both must be adjusted. This could be obtained by a modification of the input $s'$; but one would have to check also whether a change of the input gluon distribution would lead to a better agreement without coming into conflict with other measurements. Unfortunately the cross section (1) could be measured only in a rather limited range of $y$ values. In the CCFR analysis $\langle y \rangle \simeq 0.47$. Therefore it makes sense to compare an effective structure function $F_2^{eff}$ for this fixed value of $y$. As in [11] we introduce

$$F_2^{eff}(\bar{x}, Q^2, y) = (1-y)F_2^{sc}(\bar{x}, Q^2) + y^2 x F_1^{sc}(\bar{x}, Q^2) + y^2 x F_3^{sc}(\bar{x}, Q^2), \tag{15}$$

which is just the expression in the curly brackets in (1), and evaluate it for $y = 0.5$ in the two schemes. The result is plotted in Fig. 7. Comparing the results we see that the $F_2^{eff}$ are not so different. The differences are below 20%. Below $x = 0.03$ $F_2^{eff}$ in the GRV scheme is larger by up to 15% and for $x > 0.03$ it is smaller by up to 20%. Since $xF_1^{sc}$ and $F_2^{sc}$ on one side and $xF_3^{sc}$ on the other side influence $F_2^{eff}$ in opposite directions these differences cancel each other to a large extent (see Figs. 3, 4, 5). Whereas, for example at $x = 0.01$, the ratios $r$ of the results for the GRV and CTEQ3M parametrizations for $F_2^{sc}$ and $xF_3^{sc}$ are $r = 2.3$ and $r = 0.36$, respectively, the ratio is decreased to $r = 1.15$ in $F_2^{eff}$. One also notes that the difference in $xF_3^{sc}$ is responsible for the reduction of $F_2^{eff}$ in the GRV scheme for $x \geq 0.1$. These differences are significant even when we consider that the measurements of CCFR have an average experimental error of 10% as one can see from their results for the $s$ quark distribution. Therefore, we might expect that, if the CCFR analysis would be repeated in the fixed flavour scheme with the GRV structure functions as input, the prediction of the charm production cross section would deviate from the measured data. By how much is difficult to assess without going through the analysis, which must include all the experimental details like cuts, for example.

In the fixed flavour scheme the subtraction terms are quite different from those in the variable flavour scheme, i.e. we have very different $O(\alpha_s)$ corrections. In addition, a much smaller factorization scale was chosen in the former. Therefore also $xs(x)$ has to be different in the two schemes. From Fig. 7 it appears that $xs(x)$ in the GRV set can be adjusted in such a way that the differences in the $O(\alpha_s)$ correction terms for $F_2^{eff}$ between the two schemes can be compensated. Whether this can be done for the separate structure functions is an open problem. The figures show that $xF_1^{sc}$ and $F_2^{sc}$ are still approximately equal to the structure functions in the variable flavour scheme in the region $x > 0.1$, but differ appreciably for $x$ below 0.01, whereas $xF_3^{sc}$ differs considerably in the whole $x$ range. As an exercise we repeated the calculation in the fixed flavour scheme with $xs'(x)$ taken from MRS(A) and $xg(x)$ taken from GRV(94). The differences for $F_2^{eff}$ between the two schemes were reduced for large $x$ but remained at low $x$ and there was no improved agreement for the separate structure functions. A sensible test to distinguish the two schemes would be the separation of the three structure functions by measuring the $y$ dependence of the charm production cross section. This $y$ dependence reflects itself in $F_2^{eff}$ in the following way. For example at $y = 0.3$, the ratio $r$ is increased to 1.4 at $x = 0.01$ or, at $y = 0.7$ and $x = 0.1$, we find $r = 0.76$.

We studied the structure functions in the variable flavour scheme also with the scale $Q^2$ instead of $(2p_T^{max})^2$. With this scale $xF_1^{sc}(x, Q^2), F_2^{sc}(x, Q^2)$ and $xF_3^{sc}(x, Q^2)$ look somewhat different, but $F_2^{eff}(x, Q^2)$ changes very little over the whole range of $x > 0.01$. Thus considering this effective structure function as the only quantity measured so far, the CTEQ3M and MRS(A) parton distributions account very well for the CCFR data also with the scale $Q$ instead of $2p_T^{max}$. Results for the scale dependence of individual structure functions are presented in Figs. 8, 9, 10, 11 and 12. In these plots we show the structure functions for the MRS(A) parton distributions using the scales $\mu_R = \mu_s = f \cdot 2p_T^{max}$ with $f = 1/2$ and $f = 2$, respectively (dashed lines, results for the CTEQ parametrization are similar). The scale changes are significant only for small $x$. Here $xF_3^{sc}$ shows a somewhat stronger variation than the other structure functions $F_2^{sc}$ and $xF_1^{sc}$. In $F_2^{eff}$ at $y = 0.5$ (Fig. 12) this somewhat stronger variation is reduced again, so that the predictions are rather reliable with respect to the CCFR cross section. This agrees with similar findings in [5]. To obtain an estimate of the scale variation also for the GRV case we replaced the scale $4m^2$ by $4Q^2$. The results, contained also in Figs. 8 - 12 (dotted lines), show the same pattern: individual structure functions are slightly decreased (increased for $F_3^{sc}$), however $F_2^{eff}$ shows only small changes.

We also made the comparison in the DIS scheme. The results are essentially the same. The resulting structure functions in the GRV scheme are almost equal to those in the $\overline{MS}$ scheme since the parton distributions are changed only by the appropriate subtraction terms which are compensated by identical subtraction terms for the charm structure functions. The same applies for the MRS(A) parametrizations whereas for CTEQ3D new fits have been made and therefore the differences are slightly larger. It turns out, however, that the calculated charm structure functions $xF_1^{sc}$, $F_2^{sc}$ and $F_3^{sc}$ are again almost the same. In the DIS scheme the difference in $F_2^{eff}$ between the $N_f = 3$ and the $N_f = 4$ schemes is even smaller than it was for the $\overline{MS}$ scheme. However, the differences in $xF_1^{sc}$, $F_2^{sc}$ and $F_3^{sc}$ between the GRV and the variable flavour scheme stay essentially the same also in the DIS scheme. We checked that the two possible treatments of the subtraction in the DIS scheme as given in Eqs. (9) and (12) lead to small shifts of the results for $xF_1^{sc}$, $F_2^{sc}$ and $F_3^{sc}$ reaching a few percent only at small $x$, but the influence on $F_2^{eff}$ is negligible.

## 4 Concluding Remarks

In this article we have discussed the scheme dependence of predictions for the cross section of charm production in charged current deep inelastic neutrino scattering. We compared two different schemes, the so-called variable flavour scheme with 4 active flavours where the charm mass singularity is absorbed in the parton distribution functions, and the fixed flavour scheme with only 3 active flavours and no charm mass subtraction. This comparison could be performed in a consistent way by using corresponding parametrizations of parton distribution functions for $N_f = 4$ and for $N_f = 3$.

We have seen that the two schemes lead to largely differing results for separate structure functions, but the differences cancel to a large extent in the effective structure function $F_2^{eff}$ in the kinematical region of intermediate values of $y \simeq 0.5$ which has been measured in the CCFR experiment so far.

It is a general consensus that the scheme dependencies are not physical. This means it should be possible to adjust in either scheme the input distribution functions, in particular $xs(x)$ and $xg(x)$, in such a way that physical observables for charm production are reproduced. Taking the point of view that the structure functions with $N_f = 4$ and CTEQ3 or MRS(A) as input represent the "experimental data", it would be an interesting exercise to find strange quark and gluon distributions which would reproduce these "experimental data" in the $N_f = 3$ scheme. Of course, it would be preferable to start from real experimental data for the separate structure functions which unfortunately do not exist yet.

## Acknowledgement

We wish to thank A. Vogt for useful discussions.

# Figure Captions

**Fig. 1**  Diagrams contributing to charm production in neutrino scattering: **a)** Leading order diagram and gluon initiated diagrams. **b)** Diagrams with gluon radiation and virtual corrections.

**Fig. 2**  Cabibbo-rotated strange quark distribution $xs'(x, \mu^2)$ ($F_1^{sc,LO}(x,\mu^2)$ in leading order) for $\mu^2 = 25$ GeV$^2$ in the $\overline{MS}$ parametrizations of CTEQ3M (full line), MRS(A) (dashed line) and GRV(94) (dotted line).

**Fig. 3**  Structure function $xF_1^{sc}(x)$ for charm production in neutrino scattering in the $\overline{MS}$ scheme using the parametrizations of CTEQ3M (full line), MRS(A) (dashed line) and GRV(94) (dotted line).

**Fig. 4**  Same as Fig. 3 for the structure function $F_2^{sc}(x)$.

**Fig. 5**  Same as Fig. 3 for the structure function $xF_3^{sc}(x)$.

**Fig. 6**  Same as Fig. 3 for the structure function $F_L^{sc}(x)$.

**Fig. 7**  Same as Fig. 3 for the structure function $F_2^{eff}(x)$ (see eq. (15)) for $y = 0.5$.

**Fig. 8**  Scale dependence of the structure function $xF_1^{sc}(x)$ for charm production in neutrino scattering in the $\overline{MS}$ scheme using the parametrizations of GRV(94) with scale $\mu_R = 2m$ (upper dotted line) and $\mu_R = 2Q$ (lower dotted line) and for the parametrizations of MRS(A) (dashed lines) with scale $\mu_R = \mu_s = f \cdot 2p_T^{max}$. Upper line: $f = 1/2$, lower line: $f = 2$.

**Fig. 9**  Same as Fig. 8 for the structure function $F_2^{sc}(x)$. Upper lines: $\mu_R = 2m$ and $f = 1/2$, lower lines: $\mu_R = 2Q$ and $f = 2$.

**Fig. 10**  Same as Fig. 8 for the structure function $xF_3^{sc}(x)$. Upper lines: $\mu_R = 2Q$ and $f = 2$, lower lines: $\mu_R = 2m$ and $f = 1/2$.

**Fig. 11**  Same as Fig. 8 for the structure function $F_L^{sc}(x)$. Upper lines: $\mu_R = 2m$ (low $x$) and $f = 2$, lower lines: $\mu_R = 2Q$ (at low $x$) and $f = 1/2$.

**Fig. 12**  Same as Fig. 8 for the structure function $F_2^{eff}(x)$ for $y = 0.5$. Upper lines: $\mu_R = 2m$ and $f = 2$, lower lines: $\mu_R = 2Q$ and $f = 1/2$.

Fig. 1

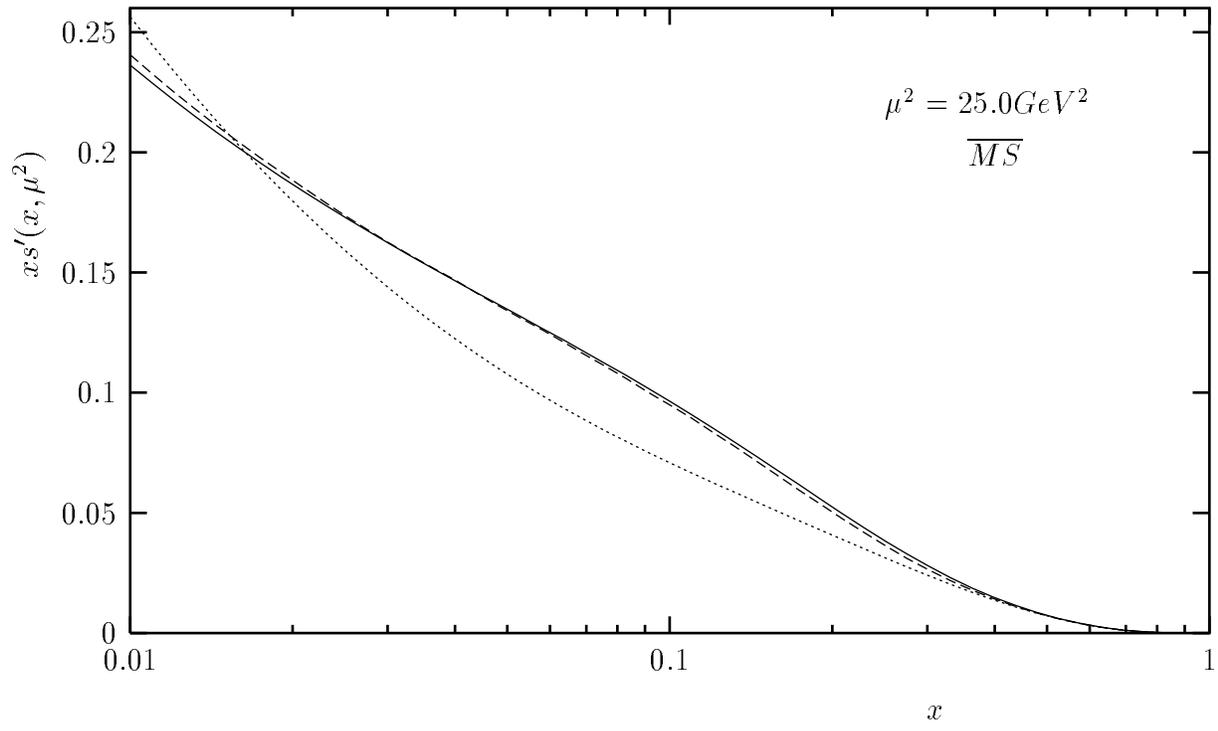

Fig. 2

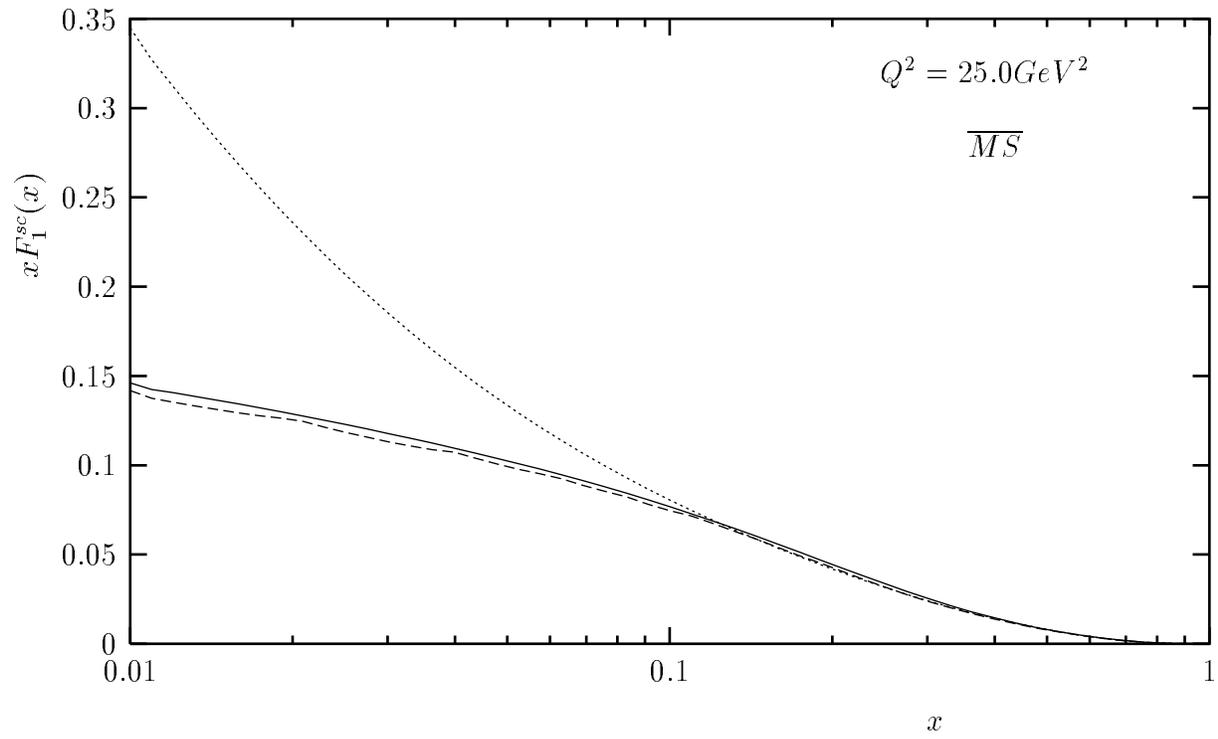

Fig. 3

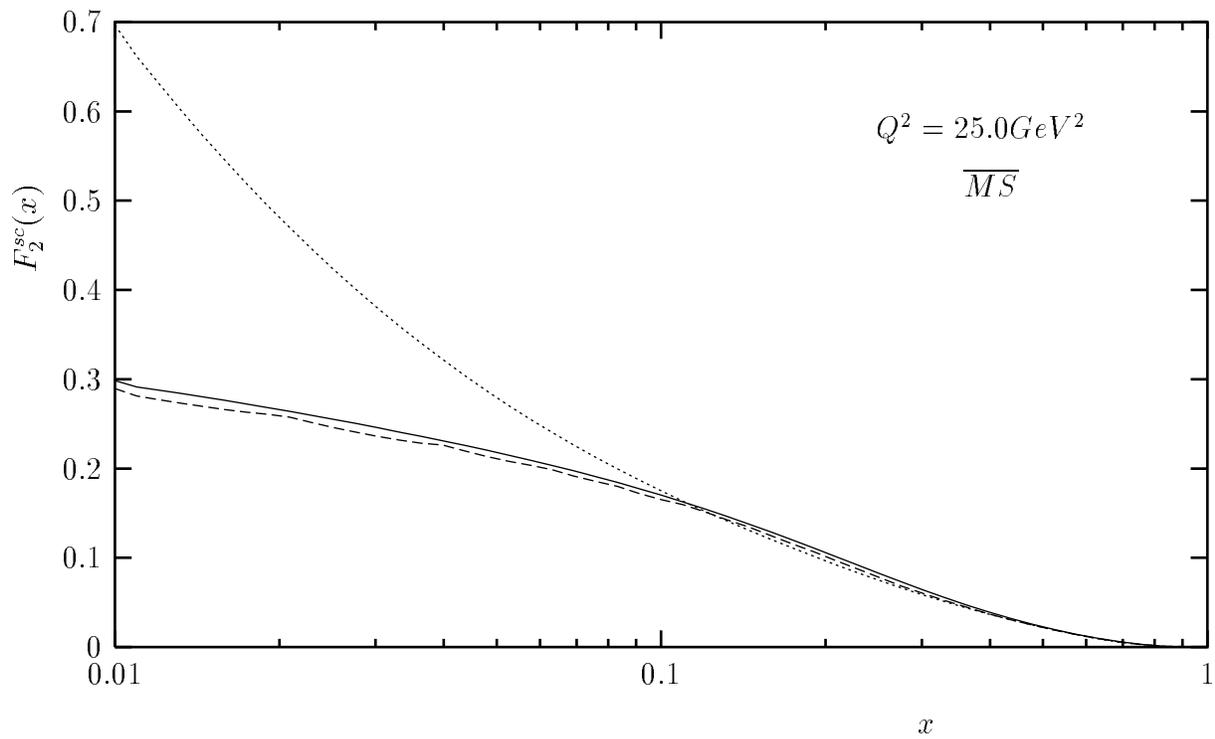

Fig. 4

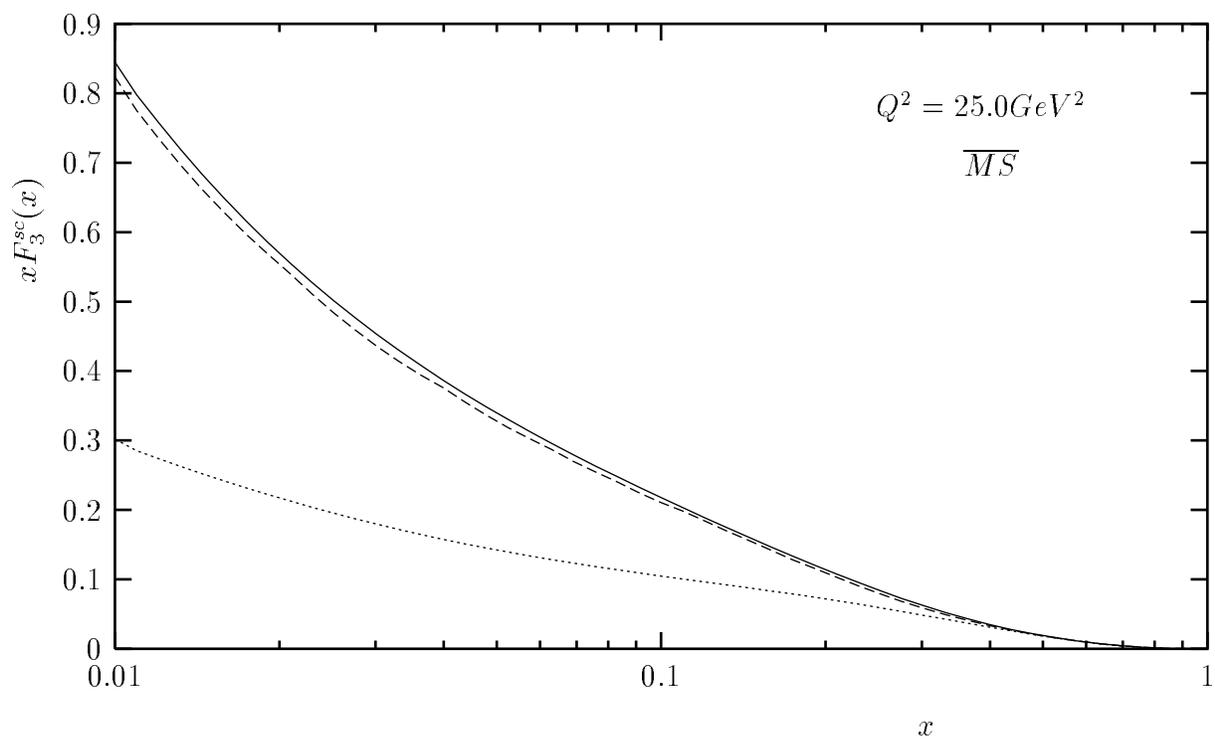

Fig. 5

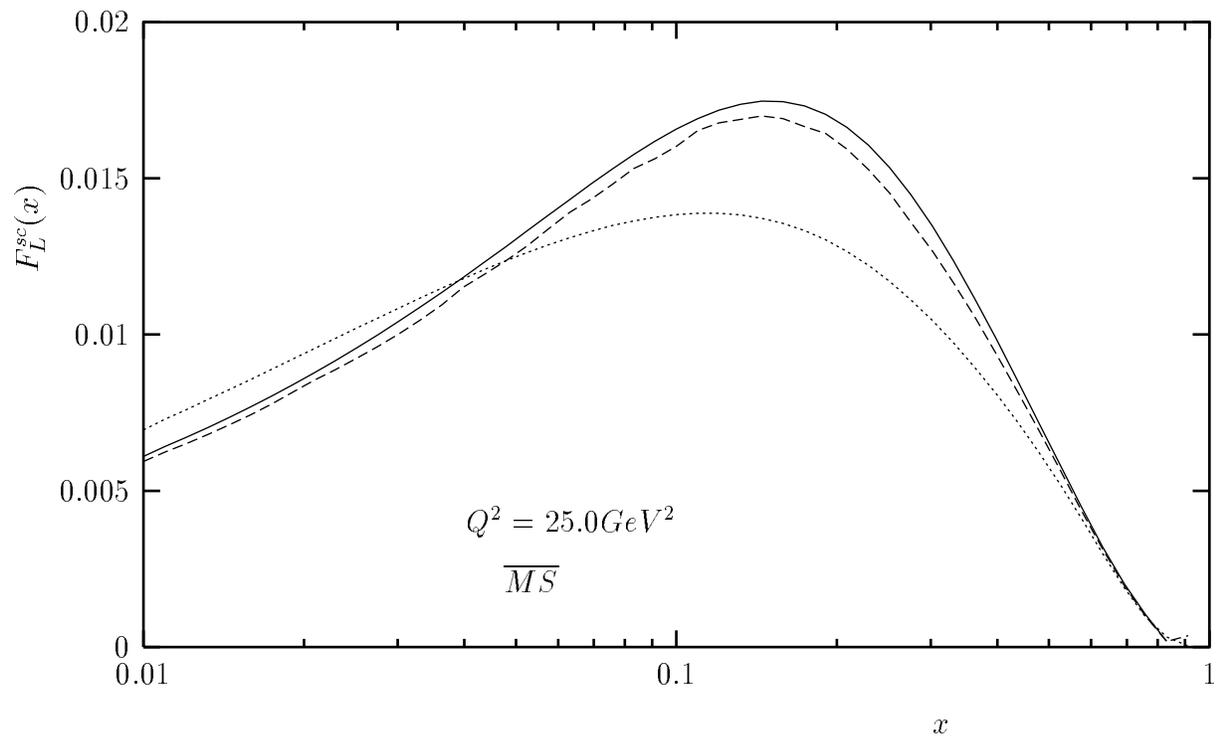

Fig. 6

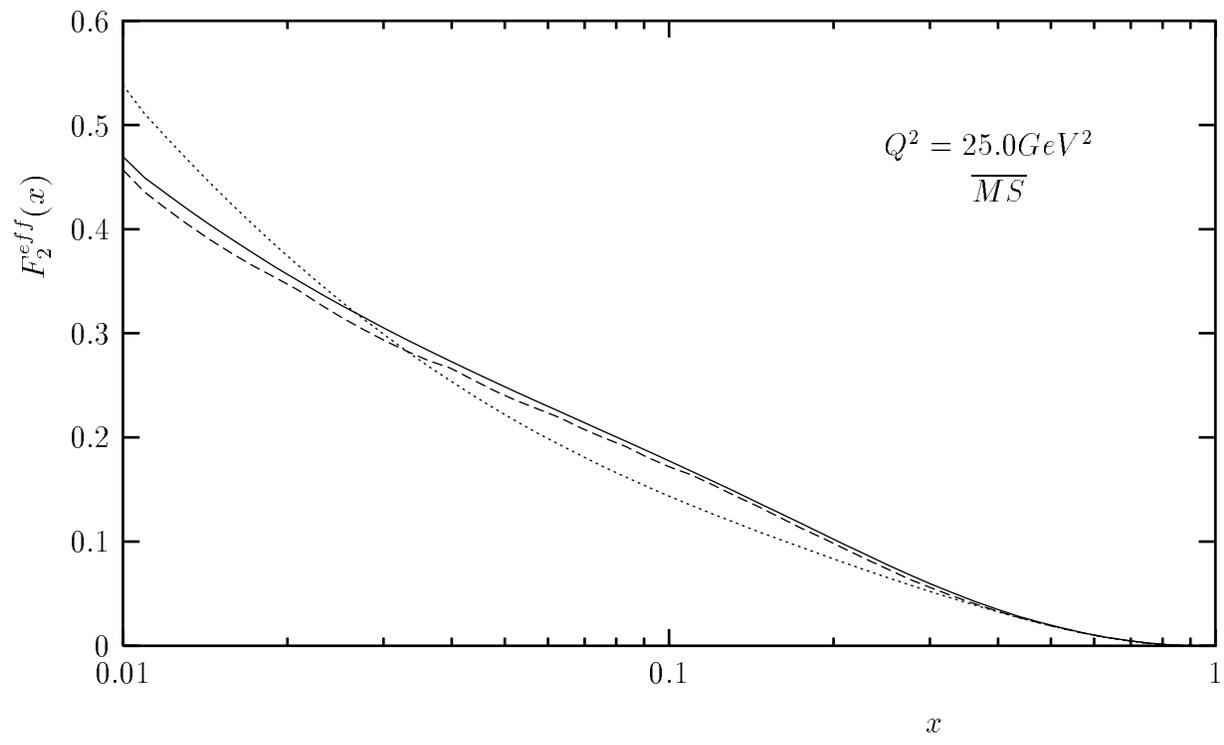

Fig. 7

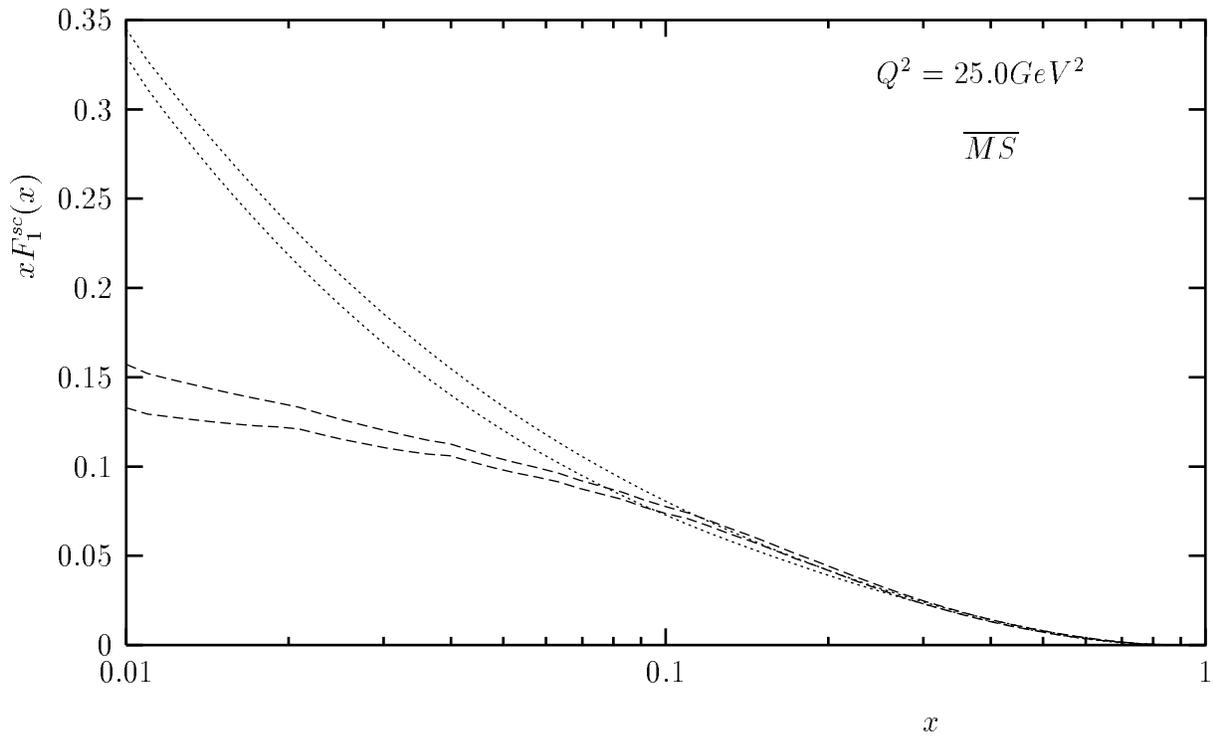

**Fig. 8**

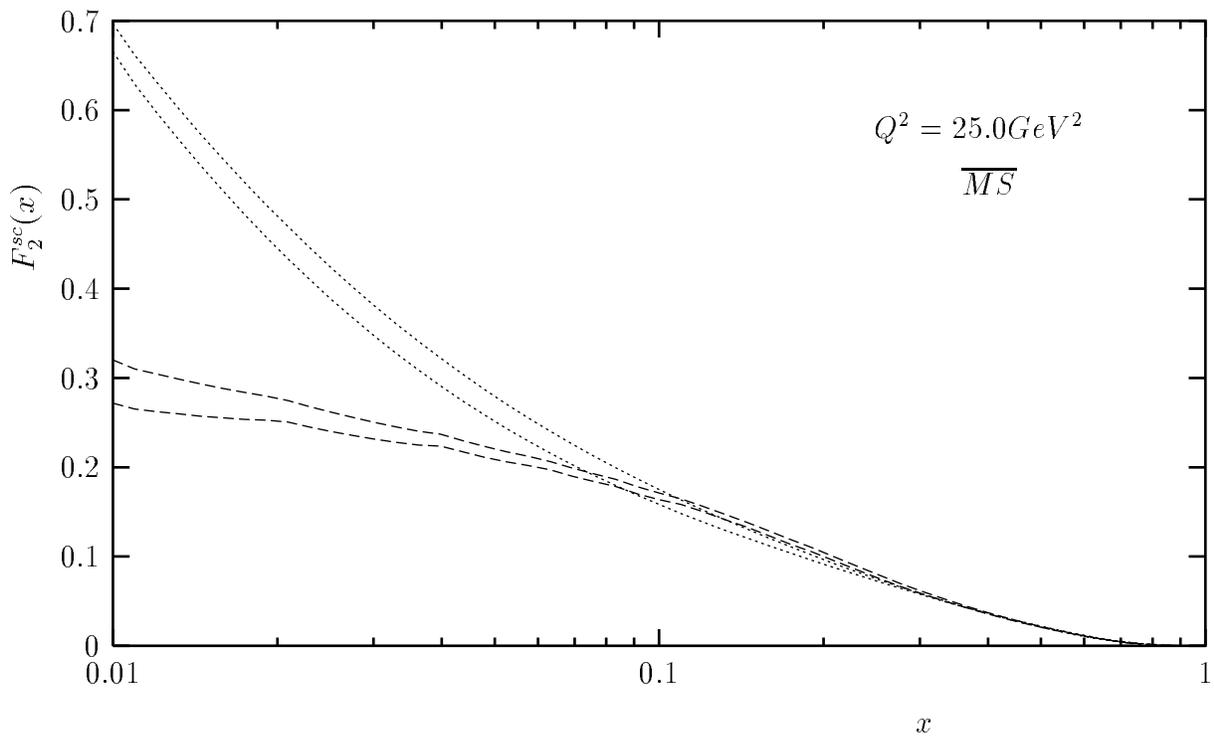

**Fig. 9**

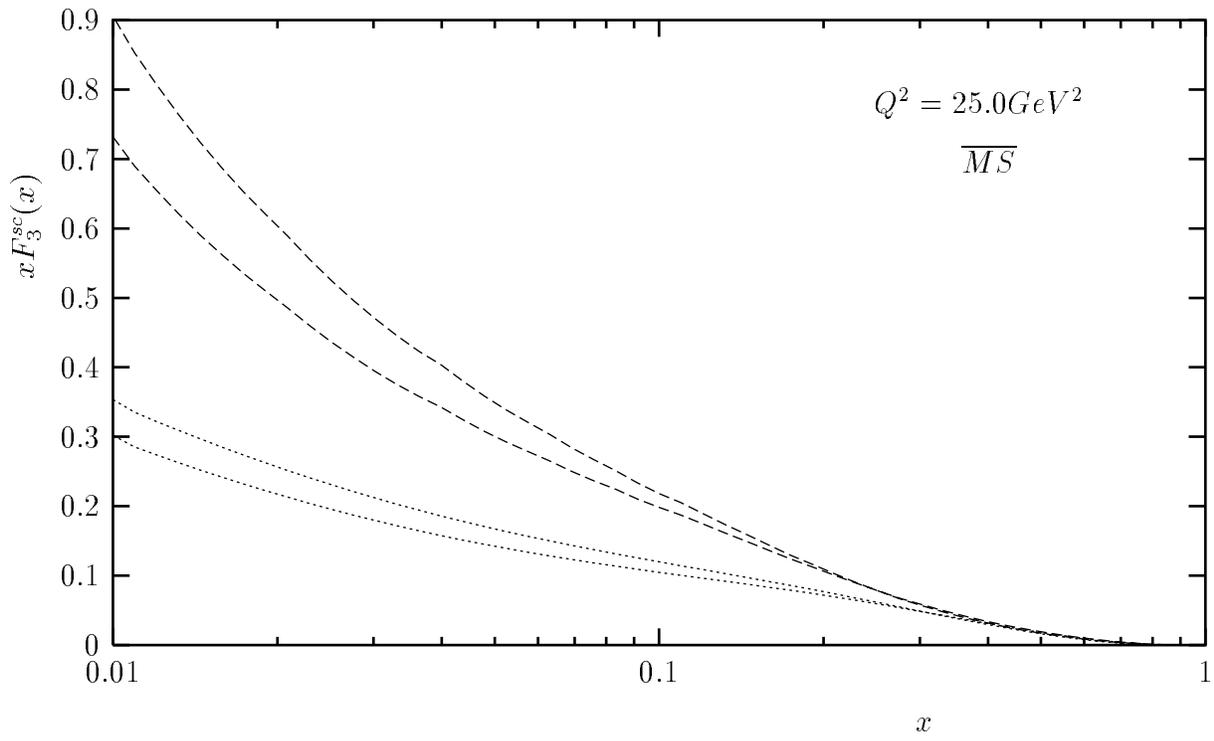

Fig. 10

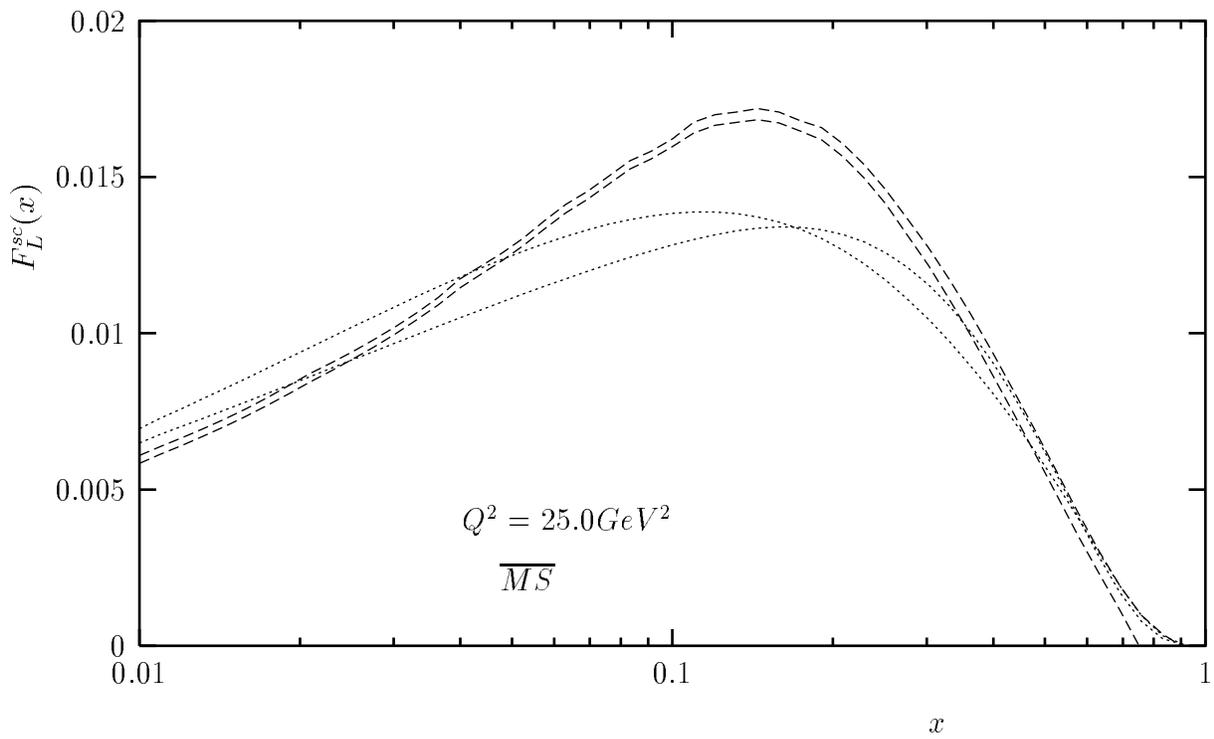

Fig. 11

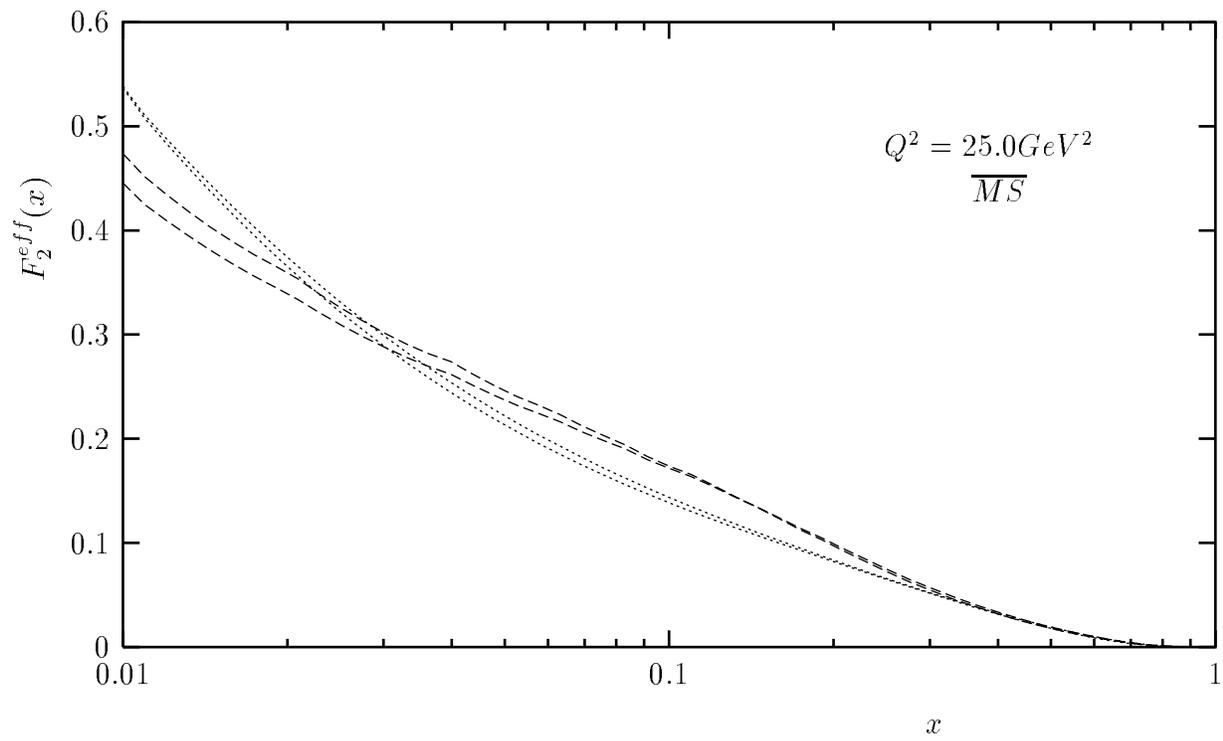

**Fig. 12**